\documentclass{llncs}

\usepackage{mathtools}
\usepackage{booktabs} 
\usepackage[parfill]{parskip} 

\begin{document}

\title{The Performance of Betting Lines for Predicting the Outcome of NFL Games}

\author{Greg Szalkowski \and Michael L. Nelson}

\institute{Old Dominion University, Department of Computer Science \\ Norfolk, VA 23529}

\maketitle      

\begin{abstract}
We investigated the performance of the collective intelligence of NFL fans predicting the outcome of games as realized through the Vegas betting lines.
Using data from 2560 games (all post-expansion, regular- and post-season games from 2002-2011), we investigated the opening and closing lines, and the margin of victory. 
We found that the line difference (the difference between the opening and closing line) could be used to retroactively predict divisional winners with no less accuracy than 75\% accuracy (i.e., ``straight up'' predictions). 
We also found that although home teams only beat the spread 47\% of the time, a strategy of betting the home team underdogs (from 2002-2011) would have produced a cumulative winning strategy of 53.5\%, above the threshold of 52.38\% needed to break even.

\end{abstract}

\section{Introduction}
Week one of the 2007 National Football League (NFL) season had the New England Patriots on the road against the New York Jets. 
The sportscasters had been talking about how the Jets were looking good this year and were ready for payback from the previous year when New England beat them in the playoffs. 
While New England had performed well in the preseason games, they had two starters out with injuries and Randy Moss was questionable for the week one game. 
The future was looking grim for the Patriots and Vegas was favoring the Jets at home by 6 points.
When betting opened, many people placed bets on New England to win even though the experts predicted that the Jets would triumph. 
The lopsided nature of the betting forced the sportsbooks to move the line repeatedly in order to keep the volume of bets even on both sides of the game. 
Eventually the line moved a total of 13 points to New England being a 7 point favorite by game day. 
New England went on to win this game 38 to 14, easily covering the spread. 
This is one example where the collective intelligence of the NFL fans was confident that New England would win even when the ``experts" thought otherwise. 

Collective intelligence is a way of synthesizing information from a group of people that no one person would have known on their own. 
Collating information from many individuals and making new conclusions based on the assimilated knowledge is the core of collective intelligence. 
Harnessing this collective intelligence using network applications is one of the fundamental ideas underlying Web 2.0 \cite{web2}. 
The archetypical example of collective intelligence is the ``guess how many jelly beans in a jar" prediction. 
Ask a group of people to estimate the number of beans in a jar, and many of the answers will be far from correct. 
But their average answer will be quite close to the real answer and maybe even closer than the prediction of most of the people who guessed.
``Under the right circumstances, groups are remarkably intelligent, and are often smarter than the smartest people in them''. 
Surowiecki \cite{surowiecki2005wisdom} outlines four conditions that must hold for collective intelligence:

\begin{quotation}There are four key qualities that make a crowd smart. It needs to be diverse, so that people are bringing different pieces of information to the table. It needs to be decentralized, so that no one at the top is dictating the crowd’s answer. It needs a way of summarizing people’s opinions into one collective verdict. And the people in the crowd need to be independent, so that they pay attention mostly to their own information, and not worrying about what everyone around them thinks. 
\end{quotation}

The NFL betting line implements each of the four qualities that Surowiecki mentions as requirements for a crowd to be smart. 
NFL fans come from all walks of life and while focused in the United States, they are a quite diverse group of people. 
The line is initially set by sportsbooks but it then moves from its initial point due to forces applied by betting volume and it neatly summarizes the views of the betting pool. 
The betting line itself could be considered a form of feedback so each of the individual bettors is not completely independent but each person is acting in their own self interest. 
This research is focused on leveraging the collective intelligence of the football community realized in the NFL point spreads and how they change as a result of the bets placed on the outcome of NFL games placed by many people \cite{surowiecki2005wisdom}.

Our question is: can the collective intelligence of the NFL fans consistently be leveraged to predict the outcome of future games?

\section{The Point Spread}

Frequently teams have a talent discrepancy that makes predicting the outcome easy; ``straight up" is the term used for simply predicting the outcome of the game and a ``point spread" is used to handicap the prediction process. 
The point spread by itself is not intended to predict a winner but to ensure that the sportsbooks make money. 
The sportsbooks make money by charging a commission, sometimes called the \textit{vigorish}, on each bet placed. 
The sportsbook's intentions are to separate the betting population in half in order to minimize risk and maximize profit. 
In order to split the population of bettors in half, the point spread on a particular game may have to be adjusted due to betting pressure on one side or the other. 
This way, the bookmaker can guarantee 5\% profit regardless of the outcome of the game (10\% from the losing half of the betting population).

Every week of the NFL season the initial point spreads are established at the beginning of the week by a small group of sportsbooks and consultants. 
Las Vegas Sports Consultants \cite{lvsc} bills itself as a provider of sports betting lines to most of the Las Vegas casinos. 
The initial point spreads are initially opened for bidding to a group of high level knowledgeable bettors and the line is adjusted based on the bets placed by those bettors \cite{harville1980predictions}. 
This initial line is sometimes referred to as the ``virgin" line and it moves to equalize the early betting so that the amount of money bet on either side of each game is roughly equal. 
After this initial adjustment the point spread is released to the public and opened for betting as the opening line.

A bettor can place a bet on either team. 
The bettor wins if the bet is on the favorite and the favorite wins by more than the point spread, or if the bet is on the predicted loser and the predicted loser either wins or loses by fewer points than the point spread. 
If the predicted winner wins by a margin equal to the point spread, a tie is declared, and the bettor neither wins or loses. 
This is usually referred to as a ``push". 
In all other cases, the bettor loses. As bets are placed the point spread values will move to continue to keep an equal volume of betting on both sides of each game.

As an example of the betting process, in the introduction New England was a 6 point underdog to the Jets in the opening line. 
The point spread was New England +6 or New York -6. 
In order to place a bet, a potential bettor must be willing to risk \$11 to win \$10. When betting opened, many people started betting on New England to win, so the sportsbooks were forced to move the line toward New England in order to entice people to bet on New York so that there would be an even betting volume on both sides. 
By taking an equal volume of bets on each side, the sportsbook is in effect charging the players betting the losing side an extra dollar.
When the game is over and the closing line was New England favored by 7 points, a person who bet \$10 on New England would receive \$21 (the \$11 risked plus the \$10 won) while the player who bet on New York loses \$11. 
The sportsbook then makes \$1 on this transaction, the \$11 lost by the New York bettor less the \$10 won by the New England bettor. 

If a \$100 bet is placed (\$110 with the ante) and wins, the bettor is paid \$100 and keeps the ante; if the bettor loses she pays \$110. 
In order to break even the bettors must win at least 52.38\% of their bets. This can be found solving equation \eqref{eq:winpercent}. 
Let the win ratio (WR) be the proportion of winning games, therefore 1-WR is the number of losing games. 
In order to determine the proportion of winning bets, WR, necessary to break even on wagering, set the expected winnings equal to the expected losses and solve for WR. 
Solving yields WR=0.5238 or 52.38\% to break even.

\begin{eqnarray}
100*WR = 110(1-WR) \\
\label{eq:winpercent} \nonumber
\end{eqnarray}

Every sportsbook that takes bets has the same incentive to maintain an equal volume of bets on each side of a given game. 
The line offered by each sportsbook is unique to that sportsbook and they can change independently of other sportsbooks. 
There is an entire industry related to ``line shopping'', that seeks to exploit differences in lines offered by different sportsbooks. 
If the line for a certain game is trending down, the ``line shoppers'' attempt to place bets at sportsbooks that move a little slower and have not adjusted down as quickly as other. 
Overall the entire ``sportsbook system'' has a form of collective intelligence feedback loop that keeps most of the lines offered by all of the sportsbooks very close if not equal to each other. 
In comparing against the spread records, there may be some differences when using line values from different sources. 
The majority of line values used in this study were obtained from \textit{The GoldSheet} \cite{goldsheet}.

\begin{figure}[h!]
	\begin{center}
		\includegraphics[width=0.8\textwidth]{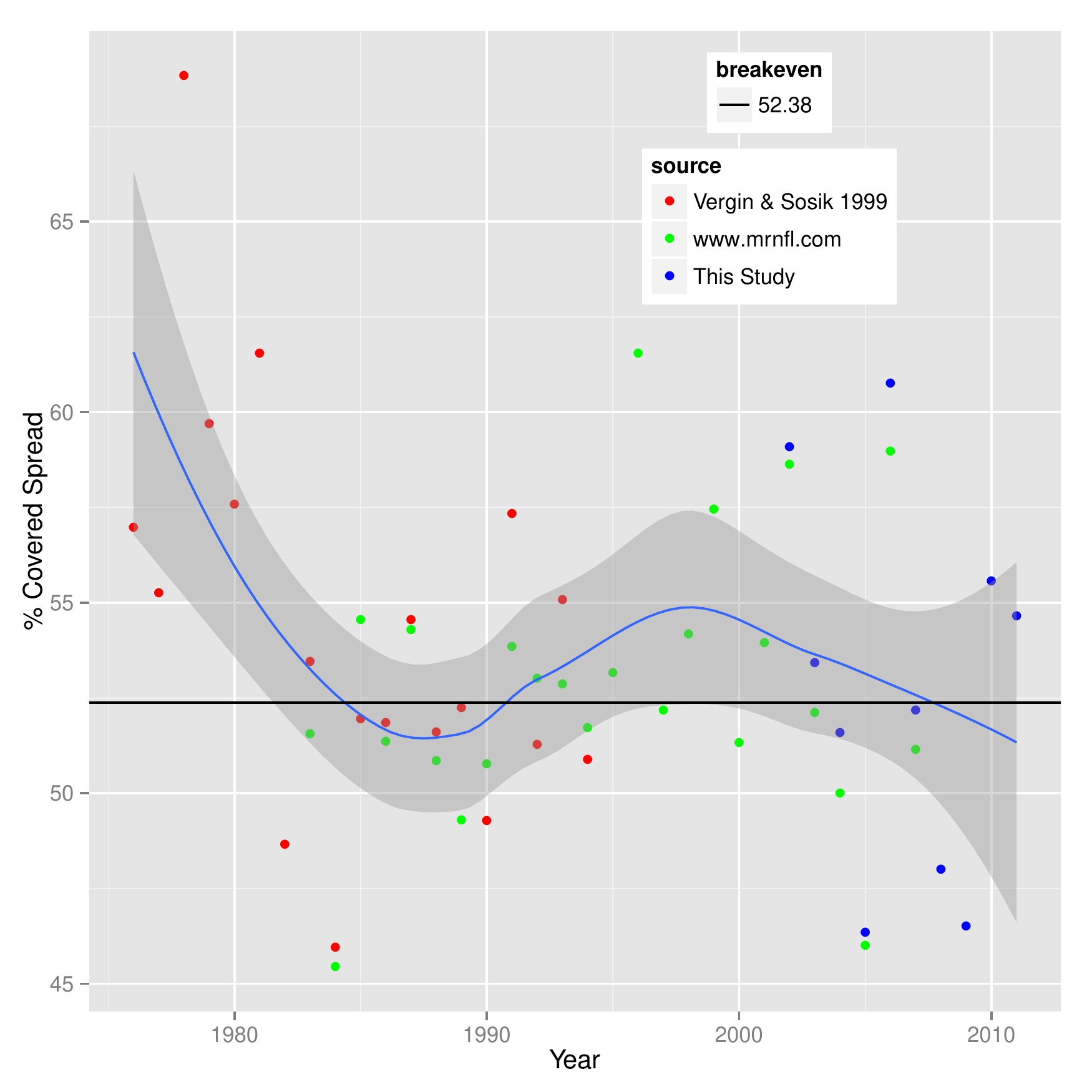}
		\caption{Home Underdogs That Covered the Point Spread}
		\label{fig:home_underdog}
	\end{center}
\end{figure}

A significant factor in determining the line is the home field advantage. 
Home teams typically have an advantage because of familiarity, travel requirements, and factors related to the crowd \cite{schwartz1977home}. 
The magnitude of this advantage is typically realized in the betting market by a three point benefit for the home team. 
At first glance the betting line appears to be a good indicator of performance, as home teams have won 58\% of the NFL games from 1981 to 1996 \cite{vergin1999no} and 57\% from 2002 to 2011. 
However winning against the spread was found to be much less decisive with the home team beating the spread in only 48.9\% of the games from 2002 to 2011 and 49.9\% of the games from 1981 to 1996 \cite{vergin1999no}. 
Home team performance against the spread for 2002 to 2011 is detailed in Table \ref{tab:homedata}. 
Historically it appears that the home game factors may have been overcompensated as the performance is below 50\%. 
One data point that sticks out is the performance of the home underdog. 
The home-underdog effect has been the subject of some research and has been attributed to late season biases \cite{borghesi2007late} and weather conditions \cite{borghesi2007home}. 
Betting on the home team underdog from 1973 to 1979 would have resulted in a 58.1\% win ratio \cite{golec1991degree}, a 52.5\% win ratio from 1981 to 1996 \cite{vergin1999no} and 53.5\% from 2002 to 2011. 
Results from a number of studies indicate that the home-underdog bias has been diminishing over the years \cite{golec1991degree,gray2012testing}. 
This diminishing bias is quite evident when plotted over time as seen in Figure \ref{fig:home_underdog}.

\begin{table}[h]
  \centering
  \caption{Home Team Records Against the Spread}
    \begin{tabular}{rrrrrrrrrrrrr}
    \toprule
          & \multicolumn{3}{c}{Favorites} & \multicolumn{3}{c}{Underdogs} & \multicolumn{3}{c}{Pick-ems} & \multicolumn{3}{c}{All Home Games} \\
    			\cmidrule(r){2-4}					\cmidrule(r){5-7} 			\cmidrule(r){8-10}    	\cmidrule(r){11-13}
          & Win   & Lose  & WR    & Win   & Lose  & WR    & Win   & Lose  & WR    & Win   & Lose  & WR \\
		\midrule
    2002  & 80    & 93    & 0.462 & 51    & 36    & 0.586 & 1     & 1     & 0.5   & 132   & 130   & 0.504 \\
    2003  & 90    & 86    & 0.511 & 36    & 35    & 0.507 & 2     & 2     & 0.5   & 128   & 123   & 0.510 \\
    2004  & 77    & 94    & 0.450 & 40    & 40    & 0.500 & 1     & 4     & 0.2   & 118   & 138   & 0.461 \\
    2005  & 91    & 76    & 0.545 & 31    & 45    & 0.408 & 1     & 0     & 1     & 123   & 121   & 0.504 \\
    2006  & 74    & 101   & 0.423 & 45    & 32    & 0.584 & 3     & 1     & 0.75  & 122   & 134   & 0.477 \\
    2007  & 82    & 79    & 0.509 & 46    & 44    & 0.511 & 0     & 1     & 0     & 128   & 124   & 0.508 \\
    2008  & 81    & 96    & 0.458 & 35    & 41    & 0.461 & 2     & 0     & 1     & 118   & 137   & 0.463 \\
    2009  & 78    & 84    & 0.481 & 39    & 46    & 0.459 & 3     & 2     & 0.6   & 120   & 132   & 0.476 \\
    2010  & 80    & 86    & 0.482 & 41    & 38    & 0.519 & 2     & 2     & 0.5   & 123   & 126   & 0.494 \\
    2011  & 83    & 93    & 0.472 & 45    & 39    & 0.536 & 0     & 0     & 0     & 128   & 132   & 0.492 \\
    Total & 816   & 888   & 0.479 & 409   & 396   & 0.508 & 15    & 13    & 0.535 & 1240  & 1297  & 0.489 \\
          &       &       &       &       &       &       &       &       &       &       &       &  \\
    0.500 &       & z=    & -1.858 &       & z=      & 0.404 &       &  z=     & 0.044 &       &   z=    & -1.907 \\
    0.524 &       & z=    & -4.002 &       & z=      & -0.955 &       &  z=     & -0.165 &       &   z=    & -5.977 \\
    \bottomrule
    \end{tabular}%
  \label{tab:homedata}%
\end{table}%

Over the years a number of researchers have investigated betting strategies in relation to NFL football \cite{woodland1991effects,tryfos1984profitability}. 
Numerous market efficiency metrics have found the betting market to be efficient \cite{vergin1999no,pankoff1968}. 
A common theme in much of the research was searching for a bias in the betting line. 
When investigating the home field advantage for bias it would be useful to compare the line to two win rates, 0.5 for a straight up win and 0.5238 to cover the spread \cite{vergin1999no}. 
The data in Table \ref{tab:homedata} shows that the favorite home team lost more often than it won from 2002 to 2011 with a z of -1.85 compared to 0.5, which is similar to the -1.36 for 1981 to 1996 \cite{vergin1999no}. 
The against the spread comparison was worse with a z of -4.002 compared to 0.5238 for 2002 to 2011 and -3.66 for 1981 to 1996.

\section{Comparing the Line with Reality}

The NFL box scores and line values for all NFL games from 2002 to 2011 were collected from the Internet and inserted into a MySQL database. 
The box scores contain over 30 statistics about each game from the final score, to statistics about individual players. 
A starting year of 2002 was chosen as this was the first year after the last NFL expansion and included all of the current 32 teams. 
Data collection resulted in 2560 regular season games in the database at 256 games per year over 10 years.

\begin{figure}[h!]
	\begin{center}
		\includegraphics[width=0.8\textwidth]{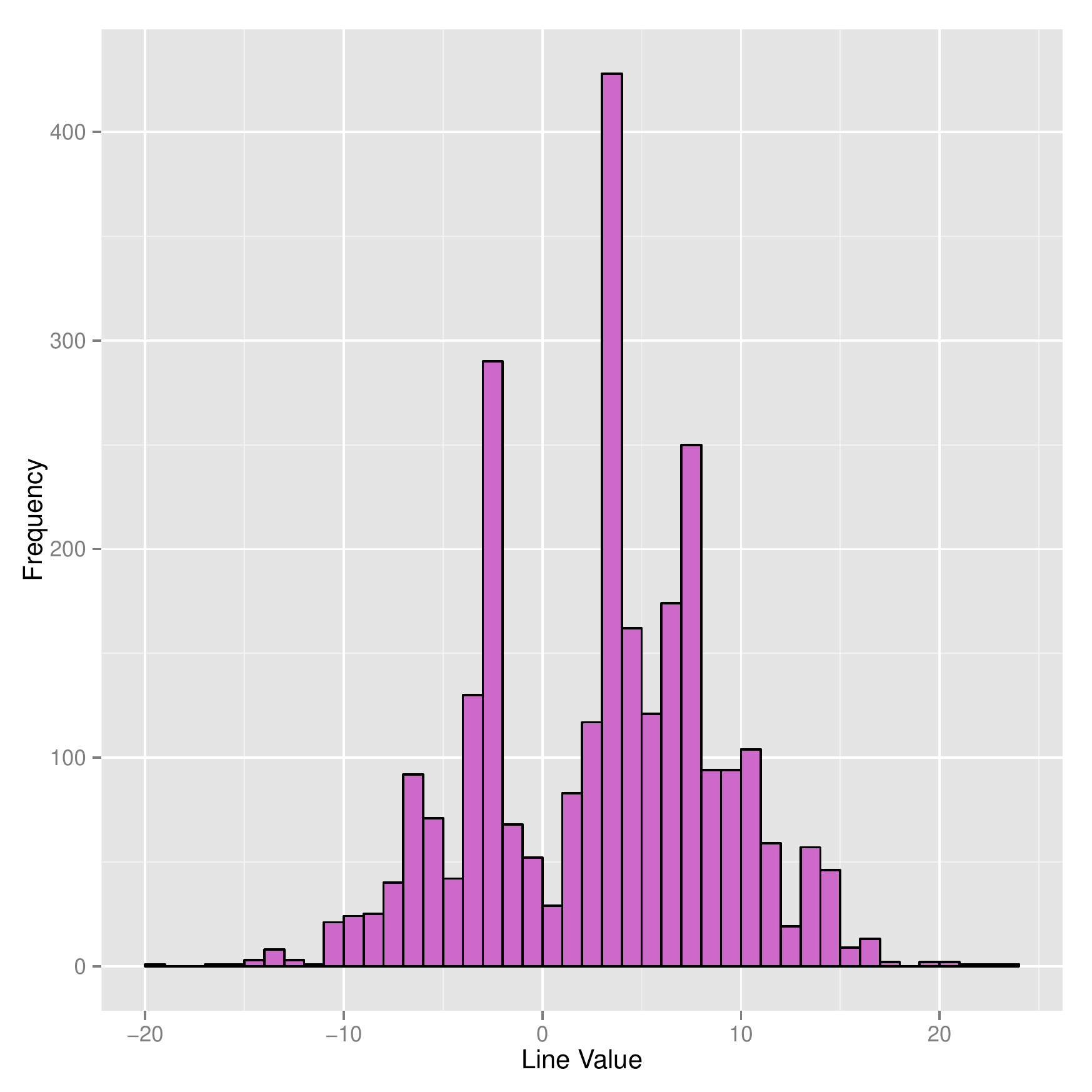}
		\caption{Most Popular Closing Line Values (2002-2011); positive values indicate the home team is favored.}
		\label{fig:line_histogram}
	\end{center}
\end{figure}

A histogram of the values of the point spreads for regular season games from 2002 to 2011 appears in Figure \ref{fig:line_histogram}. 
The most popular point spread values 3, -3, and 7 are evident and quite logical as they are common values for NFL scores \cite{nfl.com}. 
Visually there appears to be a bias towards positive values. 
A positive point spread value in this system indicates that the home team is favored, reinforcing the concept that the home team is favored more often than the visitor. 
Principal component analysis was used to dimensionally reduce the box score data and investigate the importance of each of the box score statistics. 
The betting line value had a high coefficient in almost every analysis and was ranked better than the other box score statistics with a high orthogonal variance.

To compare the point spread with actual game results, the Margin of Victory (MOV) will be compared to the point spread for each game. 
The MOV is traditionally calculated as shown in equation \eqref{eq:mov}.

\begin{equation}\label{eq:mov}
MOV = WinnerScore - LoserScore
\end{equation}

\begin{figure}[h!]
	\begin{center}
		\includegraphics[width=0.8\textwidth]{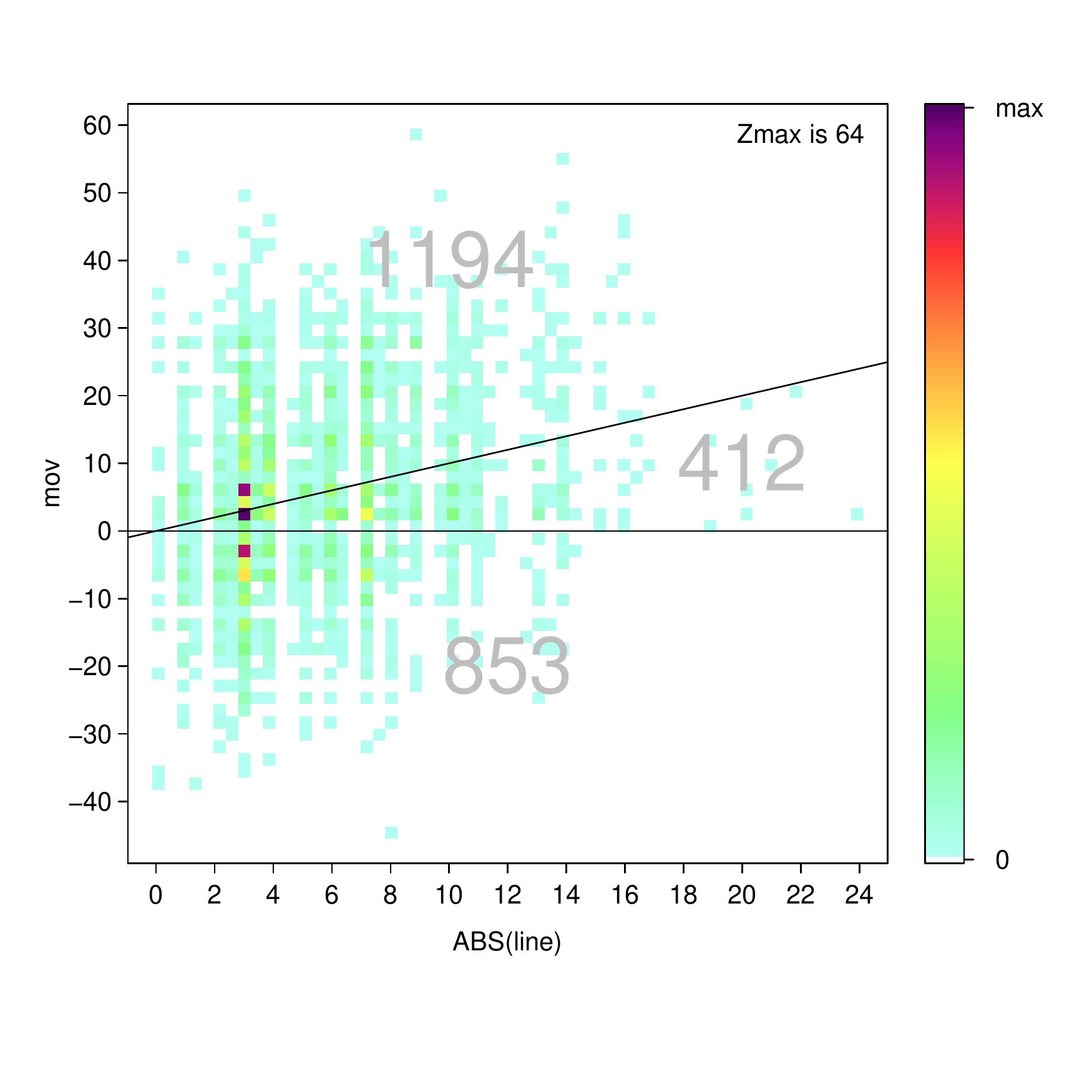}
		\caption{Closing Line vs. MOV (2002-2011); above the diagonal represents beating the spread, between the diagonal and above the x-axis is winning straight up but not beating the spread, below the x-axis is losing.}
		\label{fig:closingdiagonal}
	\end{center}
\end{figure}

The MOV and the line values can be seen in Figure \ref{fig:closingdiagonal}. 
The absolute value of the closing line is on the x-axis and the MOV on the y-axis. 
If MOV was accurate then the data in the scatter plot should be clustered about the diagonal. 
The actual results are much different and show that the line is not a good match for the actual MOV. 
Indeed the main purpose of the line is not to predict the actual MOV but to split the betting population in half. 
The games above the diagonal are games in which the favorite won. 
There were 1194 games in which the favorite beat the spread. 
The games below the diagonal but above the x-axis are games in which the favorite won but did not cover the spread. 
Games below the x-axis are games in which the favorite lost. 
There were 412 games where the favorite did not cover and 853 losses for a total of 1265 games in which the favorite lost against the spread for a loss rate of 51.5\%. 
There were 101 games which resulted in a push and are not included in the numbers on the graph. 
The line value on its own is not a good indicator of the actual MOV.

In order to overcome the deficiencies of the MOV we defined a similar metric called the Line Difference (LD) shown in equation \eqref{eq:linediff}. 
The LD for a given game is the magnitude of how far off the point spread was from the actual outcome of the game. 
A positive value will indicate that the favorite was undervalued or the underdog overvalued, and a negative value indicates that the favorite was overvalued or the underdog undervalued.

\begin{equation}\label{eq:linediff}
LD = (Favorite Score - Underdog Score) - |Closing Line|
 \end{equation}

\begin{figure}[h!]
	\begin{center}
		\includegraphics[width=0.8\textwidth]{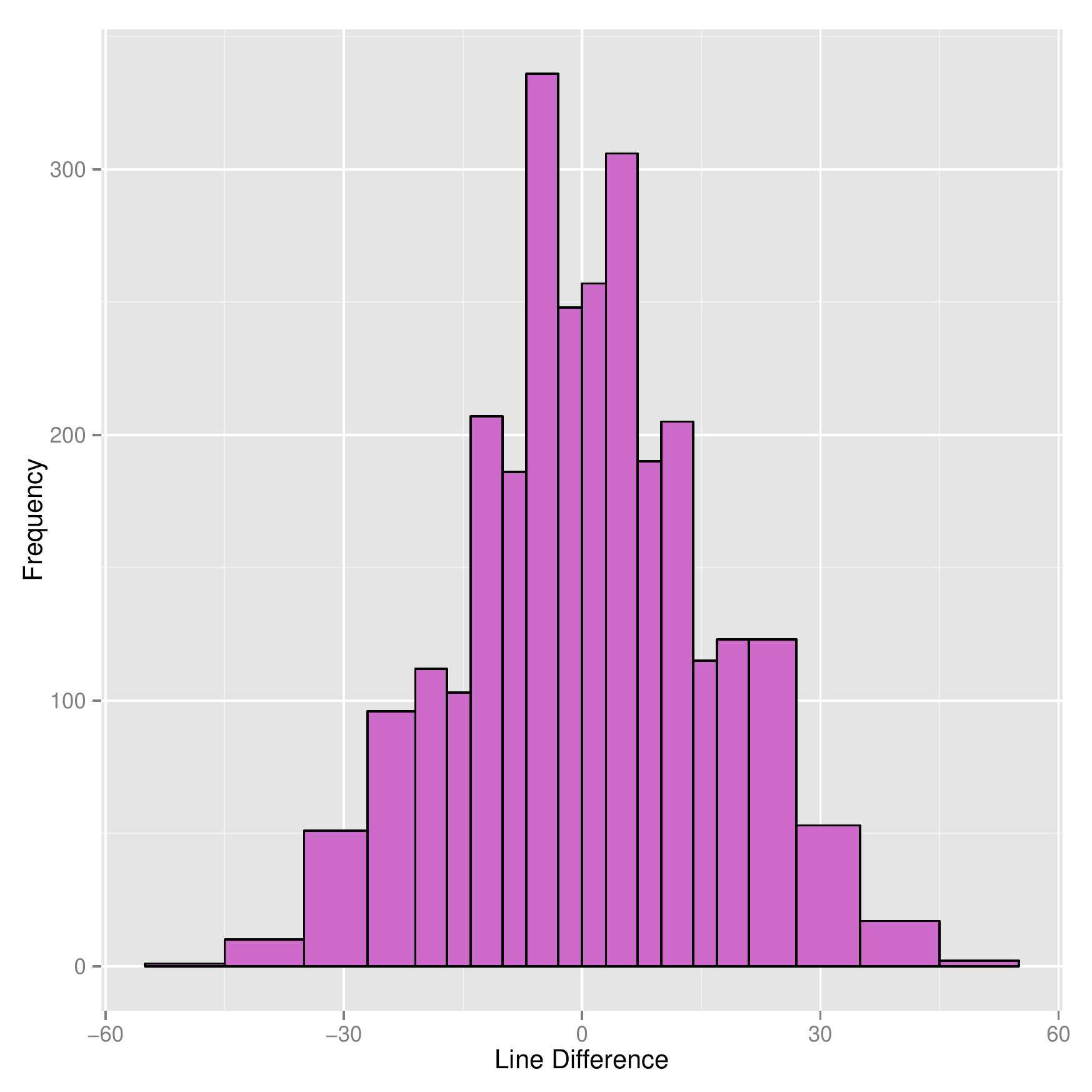}
		\caption{Histogram of Line Difference (LD) values (2002-2011)}
		\label{fig:linediff}
	\end{center}
\end{figure}	

\begin{figure}[h!]
	\begin{center}
		\includegraphics[width=0.8\textwidth]{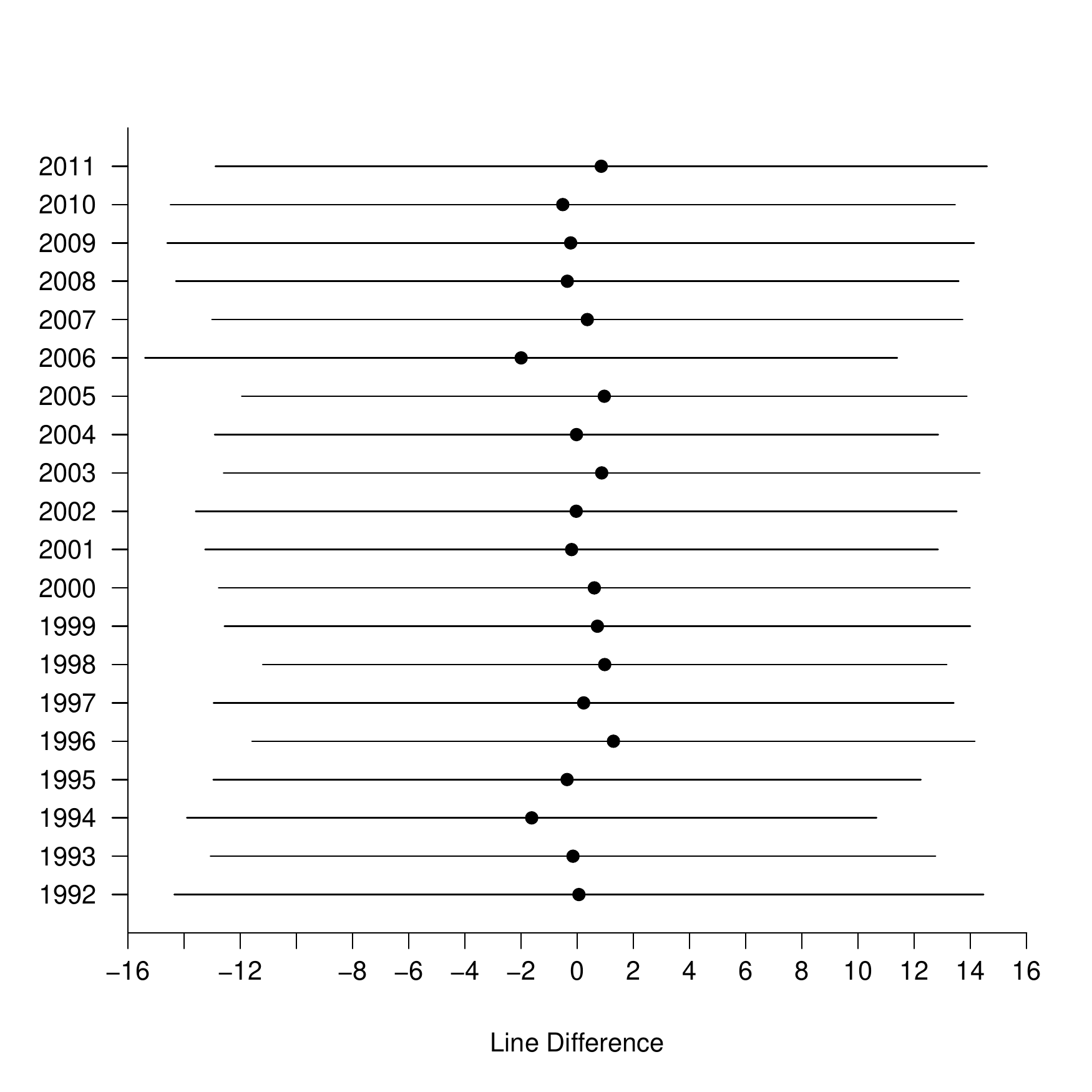}
		\caption{Mean and Deviation of Line Difference (1992-2011)}
		\label{fig:yearly_line_diff}
	\end{center}
\end{figure}

A histogram of the Line Difference values appears in Figure \ref{fig:linediff}. 
Visually the data appears close to a normal distribution with a mean of -0.009 and standard deviation on 13.588. 
This is similar to a mean of 0.07 and standard deviation 13.86 obtained from NFL seasons 1981 to 1984 \cite{stern1991probability}, 1980 to 1985 \cite{gandar1988testing}, and data for NFL seasons 1992 to 2001 \cite{nflpickles}.  
A comparison of the mean and standard deviation of the LD values from 1992 to 2011 is shown in Figure \ref{fig:yearly_line_diff}. 
The mean Line Difference form 1992 to 2011 is rather close with a value of -0.009 but the standard deviation of 13.588 which is close to two touchdowns demonstrates quite a bit of volatility in the line values.  
Additionally it appears that the line values have not improved over the years as the Line Difference values going back to at least 1980 demonstrate a rather amazing consistency.

A chi-squared goodness of fit test comparing the LD curve to a Gaussian distribution with a mean of 0 and a standard deviation of 13.588 indicates that the distribution of LD is not statistically different. 
Given that our data with a mean of 0 and standard deviation of 13.588 is an adequate approximation of a normal distribution, we can use the cumulative distribution function to approximate the probability of winning a game. 
Equation \eqref{eq:cdf} was used with p = point spread for an individual game to determine the probability that a p-point favorite would win the game \cite{stern1991probability}. 
Going back to our New England example, New England was a 7 point favorite at game time. Using equation \eqref{eq:cdf} with p=7 results in a New England win probability of 69.6\%. 
The resulting normal approximations for the probability of victory of a sample of point spreads is shown in Table \ref{tab:cumresults}.

\begin{equation}\label{eq:cdf}
Pr( F > U | P = p ) = \phi \left( \frac{p}{13.588} \right)
\end{equation}

\begin{table}[h!]
  \centering
  \caption{The Normal Approximation and the Empirical Probability of Winning}
    \begin{tabular}{ccc}
    \toprule
    Point Spread & $Pr(F > U | P)$ & Actual \\
    \midrule
    1     & 0.529 & 0.509 \\
    3     & 0.587 & 0.581 \\
    5     & 0.644 & 0.597 \\
    7     & 0.697 & 0.689 \\
    \bottomrule
    \end{tabular}%
  \label{tab:cumresults}%
\end{table}%

The probability of winning a given game could be calculated using Equation \eqref{eq:cdf} and the probability of a team winning multiple games could be envisioned as a sequence of events and calculated as the product of each of the individual events. 
For example, New England is favored by 7 points in one game and 4 points in another.
The probability of New England winning both games is $\phi \left.( \frac{7}{13.588})\right.\cdot\phi\left.( \frac{4}{13.588})\right. = 0.429 $. 
The probability of winning \textit{k} games in a season can be found by adding the the probabilities for all $\binom{16}{k}$ sequences of outcomes. 
The probabilities for all games in each season from 2002 to 2011 were calculated and each season was simulated 1000 times and the results averaged. 
The results can be found in Tables \ref{tab:2011wins} to \ref{tab:2002wins}. 
The estimates from the normal approximation are consistent with the results obtained from the actual games. 
The predicted division winners for each season were calculated and compared to the actual outcome. 
The rules for division winners become detailed in the event of ties. When there is a tie in the predicted division winner, the predicted outcome was decided in our favor. 
The results of the predictions are in Table \ref{tab:divwin}.

\begin{table}[h!]
  \centering
  \caption{Predicting the Division Winners}
    \begin{tabular}{cc}
    \toprule
    Year & Correct \\
    \midrule
    2002 & 7/8 \\
		2003 & 7/8 \\
    2004 & 6/8 \\
    2005 & 8/8 \\
		2006 & 6/8 \\
    2007 & 7/8 \\
    2008 & 7/8 \\
    2009 & 7/8 \\
		2010 & 6/8 \\
    2011 & 6/8 \\
    \bottomrule
    \end{tabular}%
  \label{tab:divwin}%
\end{table}%

Because we used the point spreads for the entire regular season, this is essentially a retrospective analysis and does not in isolation lend itself to being a good predictor of an entire season. 
In the 1970s a study of over 1,800 horse races and the accompanying betting odds was conducted. 
The winning frequency of every horse in every race was compared to the odds in the betting market. 
With only a few exceptions the betting odds precisely predicted the actual order in which the horses finished \cite{hoerl1974reliability}. 
Both the NFL and the horse races were retrospective but they reinforce the concept that the line (or odds) would be a valid input into a prediction algorithm.

\section{Line Movement}

In football, sports bettors have a whole week to bet on games. 
This is a week in which new information becomes available daily. 
Sports Illustrated puts a team on their cover\footnote{The Sports Illustrated cover jinx is an urban legend that states that individuals or teams who appear on the cover of the Sports Illustrated magazine will subsequently be jinxed (experience bad luck).}, injury reports are released, weather reports become more accurate, and paid handicappers release their picks to the public. 
All of these things have an influence on the volume and direction of betting, which directly impacts the direction and magnitude of line movement. 

Line movement is essential to sports betting. 
It is a game that begins immediately after the line makers create the virgin line. 
During early betting the sportsbooks will adjust the spread on these early lines until they get an equivalent volume of betting on both sides. 
Sportsbooks do not hope for, but expect, this early action to help firm the line. 
If the volume of betting is equal on both sides, the line is considered good and will not move. 
If the line is biased to one side or the other, heavier betting will occur one one side and continue until the line has reached a point that the collective intelligence indicates is a ``correct" line for that particular game. 
Once the line is firmed up, it will not move much until game time, but it still can.

Sportsbooks are typically reluctant to move the line too far due to the possibility of losing money on both sides of the bet. 
If the opening line has a team favored by 7 points and heavy betting forces the sportsbook to adjust the line to 9 points and the team wins by 8 points the sportsbook could be forced to pay out on bets on both sides and lose money. 
The most infamous example of this is the Stardust during the 1978 Superbowl \cite{woodland1991effects}. 
For Superbowl XIII the line varied between 3.5 and 4.5 with Pittsburgh favored. 
Pittsburgh won the game by 4 points resulting in a push for a majority of the bets. 
The sportsbooks do not make money off a push and they lost millions of dollars that year. 
Frequently it is observed that a spread may only move from 2 to 2.5 or from 3 to 3.5 before there is an equal amount of bets coming in on both sides. 
The half point moves help the sportsbooks to prevent having to pay out to both sides and to eliminate the probability of a push. 
If the games results in a push then the sportsbook needs to refund many of the bets and the overhead costs typically outweigh any profit from that game.  

\begin{figure}[h!]
	\begin{center}
		\includegraphics[width=0.8\textwidth]{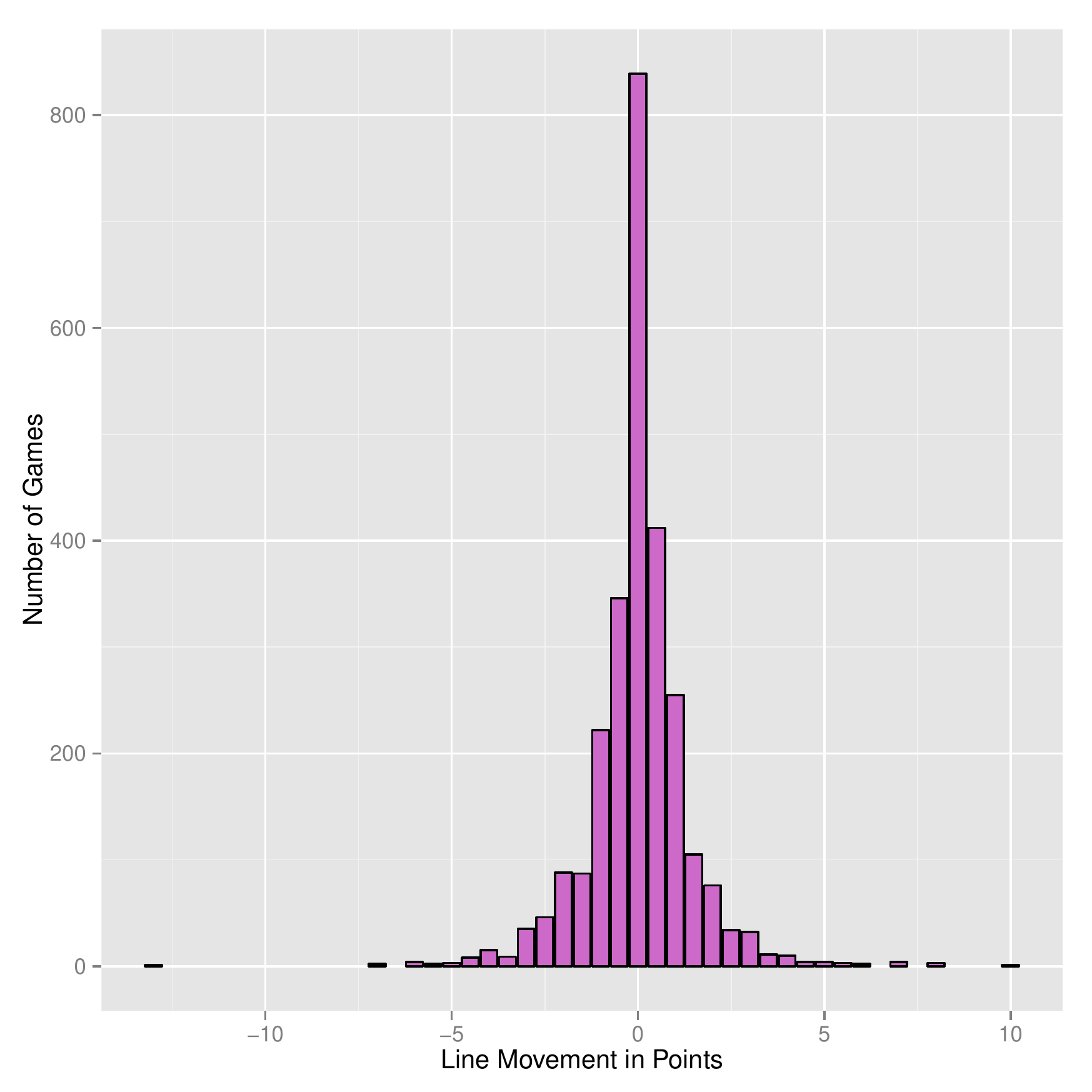}
		\caption{Line Movement in Points (2002-2012)}
		\label{fig:linemov}
	\end{center}
\end{figure}		

For the purposes of this study, the line movement is defined as the difference between opening line and the closing line. 
The line movement for each of the regular season games from 2002 to 2011 was calculated and a histogram of the values is shown in Figure \ref{fig:linemov}. 
Out of the 2560 regular season games, over 2000 games had a line movement of 1 point or less, 1548 games had a line movement of 0.5 points or less. 
Therefore only 20\% of the games had a line movement greater than one point. 
In testing the opening line versus the closing line it was found that the opening line had a smaller mean square error than the closing line but the difference between the lines was generally small and statistically insignificant \cite{gandar1988testing}. 
Subsequently all of the calculations for this study were then completed with both the opening line and the closing line. 
The results were compared and resulted in no statistically significant difference between the results. 

There is anecdotal evidence of using the temporal movement of the line for prediction of the game. 
Similar to the opening line, if sportsbooks receive more money on just one side of the spread during the week, they will adjust the line. 
This is a much slower process as bets come slowly during the week, and sportsbooks do not like to move the line once it has stabilized. 
Unless a team has a major injury or venue change, the spread should not move at this point 95\% of the time.

The last period is which the line can move is on game day. 
Big bettors that did not hammer the opening line often place their bets a few hours up to 15 minutes before game time. 
These bettors come with large amounts of money and desire the most accurate information available before the game starts. 
They utilize accurate weather reports and have access to the most current injury reports as they have delayed placing bets until the last day. 
This last minute action can be large enough to move the line right up until game time. 
While the temporal aspect of line movement may be interesting we did not have the historical data to include in our study. 
The low frequency of late week line changes would tend to lead to small sample sizes that would not have much statistical significance.

\begin{figure}[h!]
	\begin{center}
		\includegraphics[width=0.8\textwidth]{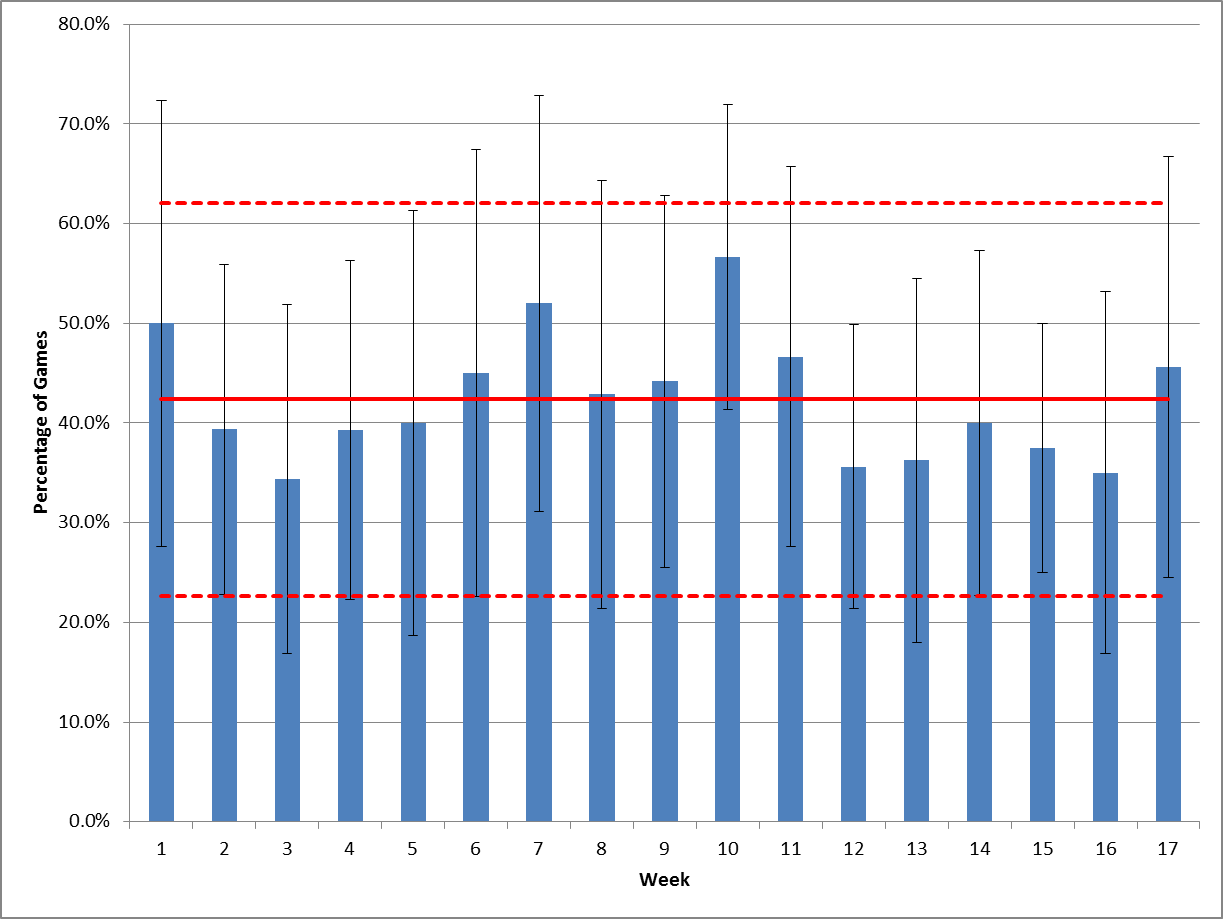}
		\caption{Percentage of Games with Line Movement $\geq 1$ By Week (2002-2011)}
		\label{fig:weeklymovement1}
	\end{center}
\end{figure}		

\begin{figure}[h!]
	\begin{center}
		\includegraphics[width=0.8\textwidth]{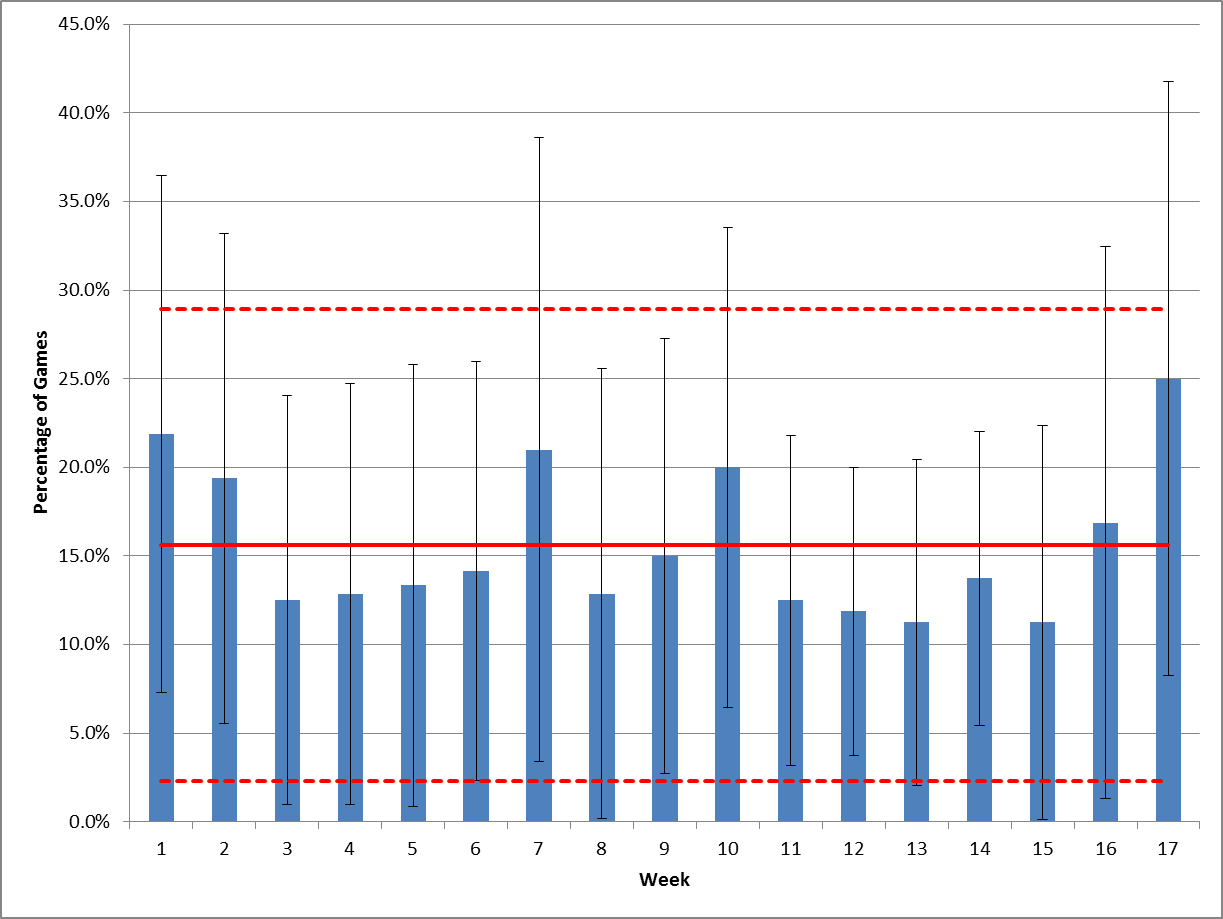}
		\caption{Percentage of Games with Line Movement $\geq 2$ By Week (2002-2011)}
		\label{fig:weeklymovement2}
	\end{center}
\end{figure}

One aspect of temporal line movement that we could analyze is to compare the amount of line movement by week to see if there is a difference. 
Because the number of games per week is not consistent throughout the year, we plotted the number of games that had a line movement greater than or equal to one point in Figure \ref{fig:weeklymovement1} and two points in Figure \ref{fig:weeklymovement2}. 
The movement was calculated as the difference between the opening and closing line. 
The values were plotted along with the mean and standard deviation in Figure \ref{fig:weeklymovement1} and Figure \ref{fig:weeklymovement2}. 
The peak in week 1 could be attributed to indeterminate performance as it is the first week of the regular season and the first time that actual game play strategies are being revealed. 
The peak in week 17 is usually attributed to teams that have locked in playoff positions who rest their starting lineup or teams that have to win to make it into the playoffs. 
With so few games demonstrating line movement of this magnitude we were unable to find a statistically significant difference that could be leveraged as a predictive value.

\section{Future Work and Conclusions}

We investigated many aspects of the betting line and its relation to NFL football. 
In comparing the opening and closing lines we found no statistically significant difference in the predictive values. 
Line movement was investigated and typically only 10\% of the games experience a difference of 2 or more points between the opening and closing line and those games did not demonstrate any consistent behavior. 
We showed that the line value using the cumulative distribution function is a good predictor of the team that will win the game straight up but historically less than 50\% accurate at predicting the winner against the spread. 

Our investigation and the Principal Component Analysis of the box score data demonstrate that the line value would be a valuable input to a machine learning algorithm for predicting the outcome of NFL games.

We also found that from 2002-2011, a strategy of betting the home underdog would produce a cumulative winning percentage of 53.5\%, which is above the threshold of 52.38\% needed to break even.  
Pre-2002 data suggests that while this strategy was historically effective, its effectivness has been reduced in later years.

Future work will include incorporating the closing line as a feature in a machine learning algorithm for predicting the outcome of games, and investigation to determine if the results here hold in college football games as well as professional (the former has more teams, but the latter has more parity between the teams).

%
%
\bibliography{nfllinebib}{}

\bibliographystyle{plain}

\appendix
\begin{table}[htbp]
  \centering
  \caption{Results of 1,000 Simulations of the 2011 NFL Season}
    \begin{tabular}{ccccc}
    \toprule
    \multicolumn{5}{c}{2011} \\
    \midrule
    Team  & Division & Predicted Wins & Actual Wins & Outcome \\
		\midrule
    New England    & AFC East & 11    & 13    & Division Winner \\
    NY Jets        & AFC East & 9     & 8     &  \\
    Buffalo        & AFC East & 7     & 6     &  \\
    Miami          & AFC East & 7     & 6     &  \\
		\\
    Baltimore      & AFC North & 11    & 12    & Division Winner \\
    Pittsburgh     & AFC North & 10    & 12    & Wild Card \\
    Cincinnati     & AFC North & 8     & 9     & Wild Card \\
    Cleveland      & AFC North & 6     & 4     &  \\
		\\
    Houston        & AFC South & 9     & 10    & Division Winner \\
    Tennessee      & AFC South & 8     & 9     &  \\
    Jacksonville   & AFC South & 6     & 5     &  \\
    Indianapolis   & AFC South & 5     & 2     &  \\
		\\
    Denver         & AFC West & 7     & 8     & Division Winner \\
    San Diego      & AFC West & 9     & 8     &  \\
    Oakland        & AFC West & 8     & 8     &  \\
    Kansas City    & AFC West & 6     & 7     &  \\
		\\
    NY Giants      & NFC East & 8     & 9     & Division Winner \\
    Dallas         & NFC East & 9     & 8     &  \\
    Philadelphia   & NFC East & 10    & 8     &  \\
    Washington     & NFC East & 7     & 5     &  \\
		\\
    Green Bay      & NFC North & 12    & 15    & Division Winner \\
    Detroit        & NFC North & 9     & 10    & Wild Card \\
    Chicago        & NFC North & 8     & 8     &  \\
    Minnesota      & NFC North & 6     & 3     &  \\
		\\
    New Orleans    & NFC South & 11    & 13    & Division Winner \\
    Atlanta        & NFC South & 9     & 10    & Wild Card \\
    Carolina       & NFC South & 7     & 6     &  \\
    Tampa Bay      & NFC South & 7     & 4     &  \\
		\\
    San Francisco    & NFC West & 9     & 13    & Division Winner \\
    Arizona   & NFC West & 7     & 8     &  \\
    Seattle   & NFC West & 6     & 7     &  \\
    St. Louis   & NFC West & 5     & 2     &  \\
    \bottomrule
    \end{tabular}%
  \label{tab:2011wins}%
\end{table}%

\begin{table}[htbp]
  \centering
  \caption{Results of 1,000 Simulations of the 2010 NFL Season}
    \begin{tabular}{ccccc}
    \toprule
    \multicolumn{5}{c}{2010} \\
    \midrule
     Team     &  Division     & Predicted Wins & Actual Wins & Outcome \\
					  \midrule
    New England & AFC East & 10    & 14    & Division Winner \\
    NY Jets & AFC East & 9     & 11    & Wild Card \\
    Miami & AFC East & 8     & 7     &  \\
    Buffalo & AFC East & 6     & 4     &  \\
		\\
    Pittsburgh & AFC North & 10    & 12    & Division Winner \\
    Baltimore & AFC North & 10    & 12    & Wild Card \\
    Cleveland & AFC North & 6     & 5     &  \\
    Cincinnati & AFC North & 7     & 4     &  \\
		\\
    Indianapolis & AFC South & 10    & 10    & Division Winner \\
    Jacksonville & AFC South & 7     & 8     &  \\
    Houston & AFC South & 8     & 6     &  \\
    Tennessee & AFC South & 8     & 6     &  \\
		\\
    Kansas City & AFC West & 8     & 10    & Division Winner \\
    San Diego & AFC West & 11    & 9     &  \\
    Oakland & AFC West & 7     & 8     &  \\
    Denver & AFC West & 7     & 4     &  \\
		\\
    Philadelphia & NFC East & 10    & 10    & Division Winner \\
    NY Giants & NFC East & 9     & 10    &  \\
    Dallas & NFC East & 8     & 6     &  \\
    Washington & NFC East & 6     & 6     &  \\
		\\
    Chicago & NFC North & 8     & 11    & Division Winner \\
    Green Bay & NFC North & 10    & 10    & Wild Card \\
    Detroit & NFC North & 6     & 6     &  \\
    Minnesota & NFC North & 8     & 6     &  \\
		\\
    Atlanta & NFC South & 10    & 13    & Division Winner \\
    New Orleans & NFC South & 11    & 11    & Wild Card \\
    Tampa Bay & NFC South & 7     & 10    &  \\
    Carolina & NFC South & 5     & 2     &  \\
		\\
    Seattle & NFC West & 8     & 7     &  Division Winner \\
    St. Louis & NFC West & 7     & 7     &  \\
    San Francisco & NFC West & 6     & 6     &  \\
    Arizona & NFC West & 7     & 5     &  \\
    \bottomrule
    \end{tabular}%
  \label{tab:2010wins}%
\end{table}%

\begin{table}[htbp]
  \centering
  \caption{Results of 1,000 Simulations of the 2009 NFL Season}
    \begin{tabular}{ccccc}
    \toprule
    \multicolumn{5}{c}{2009} \\
    \midrule
    Team  & Division & Predicted Wins & Actual Wins & Outcome \\
		 \midrule
    New England & AFC East & 10    & 10    & Division Winner \\
    NY Jets & AFC East & 9     & 9     & Wild Card \\
		Miami & AFC East & 7     & 9     &  \\
    Buffalo & AFC East & 6     & 6     &  \\
		\\
    Cincinnati & AFC North & 8     & 10    & Division Winner \\
    Baltimore & AFC North & 10    & 9     & Wild Card \\
    Pittsburgh & AFC North & 11    & 9     &  \\
    Cleveland & AFC North & 5     & 5     &  \\
		\\
    Indianapolis & AFC South & 10    & 14    & Division Winner \\
    Houston & AFC South & 9     & 9     &  \\
    Tennessee & AFC South & 8     & 8     &  \\
    Jacksonville & AFC South & 8     & 7     &  \\
		\\
    San Diego & AFC West & 10    & 13    & Division Winner \\
    Denver & AFC West & 8     & 8     &  \\
    Oakland & AFC West & 4     & 5     &  \\
    Kansas City & AFC West & 5     & 4     &  \\
		\\
    Dallas & NFC East & 10    & 11    & Division Winner \\
    Philadelphia & NFC East & 10    & 11    & Wild Card \\
    NY Giants & NFC East & 10    & 8     &  \\
    Washington & NFC East & 7     & 4     &  \\
		\\
    Minnesota & NFC North & 11    & 12    & Division Winner \\
    Green Bay & NFC North & 10    & 11    & Wild Card \\
    Chicago & NFC North & 8     & 7     &  \\
    Detroit & NFC North & 4     & 2     &  \\
		\\
    New Orleans & NFC South & 11    & 13    & Division Winner \\
    Atlanta & NFC South & 8     & 9     &  \\
    Carolina & NFC South & 7     & 8     &  \\
    Tampa Bay & NFC South & 4     & 3     &  \\
		\\
    Arizona & NFC West & 9     & 10    & Division Winner \\
    Seattle & NFC West & 8     & 8     &  \\
    San Francisco & NFC West & 7     & 5     &  \\
    St. Louis & NFC West & 4     & 1     &  \\
    \bottomrule
    \end{tabular}%
  \label{tab:2009wins}%
\end{table}%

\begin{table}[htbp]
  \centering
  \caption{Results of 1,000 Simulations of the 2008 NFL Season}
    \begin{tabular}{ccccc}
    \toprule
    \multicolumn{5}{c}{2008} \\
    \midrule
    Team  & Division & Predicted Wins & Actual Wins & Outcome \\
		\midrule
    Miami & AFC East & 8     & 11    & Division Winner \\
    New England & AFC East & 10    & 11    &  \\
    NY Jets & AFC East & 9     & 9     &  \\
    Buffalo & AFC East & 8     & 7     &  \\
		\\
    Pittsburgh & AFC North & 9     & 12    & Division Winner \\
    Baltimore & AFC North & 8     & 11    & Wild Card \\
    Cincinnati & AFC North & 5     & 4     &  \\
    Cleveland & AFC North & 6     & 4     &  \\
		\\
    Tennessee & AFC South & 10    & 13    & Division Winner \\
    Indianapolis & AFC South & 10    & 12    & Wild Card \\
    Houston & AFC South & 8     & 8     &  \\
    Jacksonville & AFC South & 8     & 5     &  \\
		\\
    San Diego & AFC West & 10    & 8     & Division Winner \\
    Denver & AFC West & 8     & 8     &  \\
    Oakland & AFC West & 5     & 5     &  \\
    Kansas City & AFC West & 5     & 2     &  \\
		\\
    NY Giants & NFC East & 10    & 12    & Division Winner \\
    Philadelphia & NFC East & 10    & 9     & Wild Card \\
		Dallas & NFC East & 10    & 9     &  \\
    Washington & NFC East & 8     & 8     &  \\
		\\
    Minnesota & NFC North & 9     & 10    & Division Winner \\
    Chicago & NFC North & 8     & 9     &  \\
    Green Bay & NFC North & 8     & 6     &  \\
    Detroit & NFC North & 4     & 0     &  \\
		\\
    Carolina & NFC South & 9     & 12    & Division Winner \\
    Atlanta & NFC South & 8     & 11    & Wild Card \\
    Tampa Bay & NFC South & 9     & 9     &  \\
    New Orleans & NFC South & 8     & 8     &  \\
		\\
    Arizona & NFC West & 9     & 9     & Division Winner \\
    Seattle & NFC West & 7     & 7     &  \\
    San Francisco & NFC West & 6     & 4     &  \\
    St. Louis & NFC West & 4     & 2     &  \\
    \bottomrule
    \end{tabular}%
  \label{tab:2008wins}%
\end{table}%

\begin{table}[htbp]
  \centering
  \caption{Results of 1,000 Simulations of the 2007 NFL Season}
    \begin{tabular}{ccccc}
    \toprule
    \multicolumn{5}{c}{2007} \\
    \midrule
    Team  & Division & Predicted Wins & Actual Wins & Outcome \\
		\midrule
    New England & AFC East & 13    & 16    & Division Winner \\
    Buffalo & AFC East & 6     & 7     &  \\
    NY Jets & AFC East & 6     & 4     &  \\
    Miami & AFC East & 5     & 1     &  \\
				\\
    Pittsburgh & AFC North & 11    & 10    & Division Winner \\
		Cleveland & AFC North & 8     & 10    &  \\
    Cincinnati & AFC North & 8     & 7     &  \\
    Baltimore & AFC North & 8     & 5     &  \\
		\\
    Indianapolis & AFC South & 11    & 13    & Division Winner \\
    Jacksonville & AFC South & 9     & 11    & Wild Card \\
    Tennessee & AFC South & 9     & 10    & Wild Card \\
    Houston & AFC South & 7     & 8     &  \\
		\\
    San Diego & AFC West & 10    & 11    & Division Winner \\
    Denver & AFC West & 8     & 7     &  \\
    Kansas City & AFC West & 6     & 4     &  \\
    Oakland & AFC West & 6     & 4     &  \\
		\\
    Dallas & NFC East & 11    & 13    & Division Winner \\
    NY Giants & NFC East & 9     & 10    & Wild Card \\
    Washington & NFC East & 8     & 9     & Wild Card \\
    Philadelphia & NFC East & 9     & 8     &  \\
		\\
    Green Bay & NFC North & 9     & 13    & Division Winner \\
    Minnesota & NFC North & 8     & 8     &  \\
    Chicago & NFC North & 8     & 7     &  \\
    Detroit & NFC North & 7     & 7     &  \\
		\\
    Tampa Bay & NFC South & 8     & 9     & Division Winner \\
    Carolina & NFC South & 7     & 7     &  \\
    New Orleans & NFC South & 9     & 7     &  \\
    Atlanta & NFC South & 6     & 4     &  \\
		\\
    Seattle & NFC West & 10    & 10    & Division Winner \\
    Arizona & NFC West & 8     & 8     &  \\
    San Francisco & NFC West & 5     & 5     &  \\
    St. Louis & NFC West & 6     & 3     &  \\
    \bottomrule
    \end{tabular}%
  \label{tab:2007wins}%
\end{table}%

\begin{table}[htbp]
  \centering
  \caption{Results of 1,000 Simulations of the 2006 NFL Season}
    \begin{tabular}{ccccc}
    \toprule
    \multicolumn{5}{c}{2006} \\
    \midrule
    Team  & Division & Predicted Wins & Actual Wins & Outcome \\
		\midrule
    New England & AFC East & 10    & 12    & Division Winner \\
    NY Jets & AFC East & 7     & 10    & Wild Card \\
    Buffalo & AFC East & 6     & 7     &  \\
    Miami & AFC East & 8     & 6     &  \\
		\\
    Baltimore & AFC North & 9     & 13    & Division Winner \\
    Cincinnati & AFC North & 9     & 8     &  \\
    Pittsburgh & AFC North & 9     & 8     &  \\
    Cleveland & AFC North & 6     & 4     &  \\
		\\
    Indianapolis & AFC South & 11    & 12    & Division Winner \\
    Jacksonville & AFC South & 9     & 8     &  \\
    Tennessee & AFC South & 5     & 8     &  \\
    Houston & AFC South & 5     & 6     &  \\
		\\
    San Diego & AFC West & 11    & 14    & Division Winner \\
    Denver & AFC West & 10    & 9     &  \\
    Kansas City & AFC West & 8     & 9     & Wild Card \\
    Oakland & AFC West & 5     & 2     &  \\
		\\
    Philadelphia & NFC East & 9     & 10    & Division Winner \\
    Dallas & NFC East & 10    & 9     & Wild Card \\
    NY Giants & NFC East & 8     & 8     & Wild Card \\
    Washington & NFC East & 7     & 5     &  \\
		\\
    Chicago & NFC North & 11    & 13    & Division Winner \\
    Green Bay & NFC North & 6     & 8     &  \\
    Minnesota & NFC North & 8     & 6     &  \\
    Detroit & NFC North & 6     & 3     &  \\
		\\
    New Orleans & NFC South & 8     & 10    & Division Winner \\
    Carolina & NFC South & 9     & 8     &  \\
    Atlanta & NFC South & 8     & 7     &  \\
    Tampa Bay & NFC South & 6     & 4     &  \\
		\\
    Seattle & NFC West & 9     & 9     & Division Winner \\
    St. Louis & NFC West & 8     & 8     &  \\
    San Francisco & NFC West & 6     & 7     &  \\
    Arizona & NFC West & 6     & 5     &  \\
    \bottomrule
    \end{tabular}%
  \label{tab:2006wins}%
\end{table}%

\begin{table}[htbp]
  \centering
  \caption{Results of 1,000 Simulations of the 2005 NFL Season}
    \begin{tabular}{ccccc}
    \toprule
    \multicolumn{5}{c}{2005}              \\
    \midrule
    Team  & Division & Predicted Wins & Actual Wins & Outcome \\
		\midrule
    New England & AFC East & 9     & 10    & Division Winner \\
    Miami & AFC East & 7     & 9     &  \\
    Buffalo & AFC East & 7     & 5     &  \\
    NY Jets & AFC East & 6     & 4     &  \\
		\\
    Cincinnati & AFC North & 10    & 11    & Division Winner \\
    Pittsburgh & AFC North & 10    & 11    & Wild Card \\
    Baltimore & AFC North & 7     & 6     &  \\
    Cleveland & AFC North & 6     & 6     &  \\
		\\
    Indianapolis & AFC South & 12    & 14    & Division Winner \\
    Jacksonville & AFC South & 9     & 12    & Wild Card \\
    Tennessee & AFC South & 6     & 4     &  \\
    Houston & AFC South & 5     & 2     &  \\
		\\
    Denver & AFC West & 10    & 13    & Division Winner \\
    Kansas City & AFC West & 8     & 10    &  \\
    San Diego & AFC West & 10    & 9     &  \\
    Oakland & AFC West & 7     & 4     &  \\
		\\
    NY Giants & NFC East & 9     & 11    & Division Winner \\
    Washington & NFC East & 9     & 10    & Wild Card \\
    Dallas & NFC East & 8     & 9     &  \\
    Philadelphia & NFC East & 8     & 6     &  \\
		\\
    Chicago & NFC North & 8     & 11    & Division Winner \\
    Minnesota & NFC North & 8     & 9     &  \\
    Detroit & NFC North & 7     & 5     &  \\
    Green Bay & NFC North & 7     & 4     &  \\
		\\
    Carolina & NFC South & 10    & 11    & Wild Card \\
    Tampa Bay & NFC South & 9     & 11    & Division Winner \\
    Atlanta & NFC South & 9     & 8     &  \\
    New Orleans & NFC South & 6     & 3     &  \\
		\\
    Seattle & NFC West & 10    & 13    & Division Winner \\
    St. Louis & NFC West & 8     & 6     &  \\
    Arizona & NFC West & 7     & 5     &  \\
    San Francisco & NFC West & 4     & 4     &  \\
    \bottomrule
    \end{tabular}%
  \label{tab:2005wins}%
\end{table}%

\begin{table}[htbp]
  \centering
  \caption{Results of 1,000 Simulations of the 2004 NFL Season}
    \begin{tabular}{ccccc}
    \toprule
    \multicolumn{5}{c}{2004}              \\
    \midrule
    Team  & Division & Predicted Wins & Actual Wins & Outcome \\
		\midrule
    New England & AFC East & 11    & 14    & Division Winner \\
    NY Jets & AFC East & 9     & 10    & Wild Card \\
    Buffalo & AFC East & 8     & 9     &  \\
    Miami & AFC East & 6     & 4     &  \\
		\\
    Pittsburgh & AFC North & 9     & 15    & Division Winner \\
    Baltimore & AFC North & 9     & 9     &  \\
    Cincinnati & AFC North & 7     & 8     &  \\
    Cleveland & AFC North & 6     & 4     &  \\
		\\
    Indianapolis & AFC South & 10    & 12    & Division Winner \\
    Jacksonville & AFC South & 8     & 9     &  \\
    Houston & AFC South & 6     & 7     &  \\
    Tennessee & AFC South & 8     & 5     &  \\
		\\
    San Diego & AFC West & 8     & 12    & Division Winner \\
    Denver & AFC West & 10    & 10    & Wild Card \\
    Kansas City & AFC West & 9     & 7     &  \\
    Oakland & AFC West & 7     & 5     &  \\
		\\
    Philadelphia & NFC East & 11    & 13    & Division Winner \\
    Washington & NFC East & 7     & 6     &  \\
    NY Giants & NFC East & 7     & 6     &  \\
    Dallas & NFC East & 7     & 6     &  \\
		\\
    Green Bay & NFC North & 9     & 10    & Division Winner \\
    Minnesota & NFC North & 10    & 8     & Wild Card \\
    Detroit & NFC North & 7     & 6     &  \\
    Chicago & NFC North & 6     & 5     &  \\
		\\
    Atlanta & NFC South & 9     & 11    & Division Winner \\
    New Orleans & NFC South & 7     & 8     &  \\
    Carolina & NFC South & 8     & 7     &  \\
    Tampa Bay & NFC South & 8     & 5     &  \\
		\\
    Seattle & NFC West & 10    & 9     & Division Winner \\
    St. Louis & NFC West & 9     & 8     & Wild Card \\
    Arizona & NFC West & 6     & 6     &  \\
    San Francisco & NFC West & 5     & 2     &  \\
    \bottomrule
    \end{tabular}%
  \label{tab:2004wins}%
\end{table}%

\begin{table}[htbp]
  \centering
  \caption{Results of 1,000 Simulations of the 2003 NFL Season}
    \begin{tabular}{ccccc}
    \toprule
		\multicolumn{5}{c}{2003}              \\
    \midrule
    Team  & Division & Predicted Wins & Actual Wins & Outcome \\
    \midrule
    New England & AFC East & 9     & 14    & Division Winner \\
    Miami & AFC East & 9     & 10    &  \\
    Buffalo & AFC East & 8     & 6     &  \\
    NY Jets & AFC East & 7     & 6     &  \\
		\\
    Baltimore & AFC North & 9     & 10    & Division Winner \\
    Cincinnati & AFC North & 7     & 8     &  \\
    Pittsburgh & AFC North & 8     & 6     &  \\
    Cleveland & AFC North & 7     & 5     &  \\
		\\
    Indianapolis & AFC South & 10    & 12    & Division Winner \\
    Tennessee & AFC South & 10    & 12    &  Wild Card\\
    Houston & AFC South & 5     & 5     &  \\
    Jacksonville & AFC South & 7     & 5     &  \\
		\\
    Kansas City & AFC West & 11    & 13    & Division Winner \\
    Denver & AFC West & 9     & 10    & Wild Card \\
    Oakland & AFC West & 7     & 4     &  \\
    San Diego & AFC West & 6     & 4     &  \\
		\\
    Philadelphia & NFC East & 9     & 12    & Division Winner \\
    Dallas & NFC East & 8     & 10    &  Wild Card\\
    Washington & NFC East & 7     & 5     &  \\
    NY Giants & NFC East & 8     & 4     &  \\
		\\
    Green Bay & NFC North & 10    & 10    & Division Winner \\
    Minnesota & NFC North & 10    & 9     &  \\
    Chicago & NFC North & 6     & 7     &  \\
    Detroit & NFC North & 5     & 5     &  \\
		\\
    Carolina & NFC South & 9     & 11    & Division Winner \\
    New Orleans & NFC South & 8     & 8     &  \\
    Tampa Bay & NFC South & 10    & 7     &  \\
    Atlanta & NFC South & 6     & 5     &  \\
		\\
    St. Louis & NFC West & 10    & 12    & Division Winner \\
    Seattle & NFC West & 9     & 10    & Wild Card \\
    San Francisco & NFC West & 8     & 7     &  \\
    Arizona & NFC West & 5     & 4     &  \\
    \bottomrule
    \end{tabular}%
  \label{tab:2003wins}%
\end{table}%

\begin{table}[htbp]
  \centering
  \caption{Results of 1,000 Simulations of the 2002 NFL Season}
    \begin{tabular}{ccccc}
    \toprule
    \multicolumn{5}{c}{2002}              \\
    \midrule
    Team  & Division & Predicted Wins & Actual Wins & Outcome \\
     \midrule
    NY Jets & AFC East & 8     & 9     & Division Winner \\
    Miami & AFC East & 9     & 9     &  \\
    New England & AFC East & 9     & 9     &  \\
    Buffalo & AFC East & 8     & 8     &  \\
		\\
    Pittsburgh & AFC North & 10    & 10    & Division Winner \\
    Cleveland & AFC North & 8     & 9     & Wild Card \\
    Baltimore & AFC North & 7     & 7     &  \\
    Cincinnati & AFC North & 6     & 2     &  \\
		\\
    Tennessee & AFC South & 9     & 11    & Division Winner \\
    Indianapolis & AFC South & 9     & 10    &  Wild Card\\
    Jacksonville & AFC South & 8     & 6     &  \\
    Houston & AFC South & 4     & 4     &  \\
		\\
    Oakland & AFC West & 10    & 11    & Division Winner \\
    Denver & AFC West & 9     & 9     &  \\
    Kansas City & AFC West & 8     & 8     &  \\
    San Diego & AFC West & 8     & 8     &  \\
		\\
    Philadelphia & NFC East & 10    & 12    & Division Winner \\
    NY Giants & NFC East & 8     & 10    & Wild Card \\
    Washington & NFC East & 7     & 7     &  \\
    Dallas & NFC East & 7     & 5     &  \\
		\\
    Green Bay & NFC North & 10    & 12    & Division Winner \\
    Minnesota & NFC North & 7     & 6     &  \\
    Chicago & NFC North & 7     & 4     &  \\
    Detroit & NFC North & 6     & 3     &  \\
		\\
    Tampa Bay & NFC South & 10    & 12    & Division Winner \\
    Atlanta & NFC South & 9     & 9     &  Wild Card\\
    New Orleans & NFC South & 9     & 9     &  \\
    Carolina & NFC South & 6     & 7     &  \\
		\\
    San Francisco & NFC West & 10    & 10    & Division Winner \\
    St. Louis & NFC West & 9     & 7     &  \\
    Seattle & NFC West & 7     & 7     &  \\
    Arizona & NFC West & 6     & 5     &  \\
    \bottomrule
    \end{tabular}%
  \label{tab:2002wins}%
\end{table}%

\end{document}